\documentclass[12pt]{iopart}
\usepackage{graphicx}
\usepackage{caption}

\begin{document}
\title{Spin-orbit torques from topological insulator surface states: Effect of extrinsic spin-orbit scattering on an out-of-plane magnetization}
\author{Mohsen Farokhnezhad}
\address{School of Nanoscience, Institute for Research in Fundamental Sciences, IPM, Tehran, 19395-5531, Iran}
\author{Reza Asgari}
\ead{r.asgari@unsw.edu.au, asgari@ipm.ir}
\address{School of Physics, Institute for Research in Fundamental Sciences, IPM, Tehran, 19395-5531, Iran}
\address{School  of  Physics,  University  of  New  South  Wales,  Kensington,  NSW  2052,  Australia}
\author{Dimitrie Culcer}
\address{School  of  Physics,  University  of  New  South  Wales,  Kensington,  NSW  2052,  Australia}
\vspace{10pt}
\begin{indented}
\item[]July 2022
\end{indented}

%
%
%
%
%

\begin{abstract}
The origins of the spin-orbit torque (SOT) at ferromagnet/topological insulator (FM/TI) interfaces are incompletely understood. Theory has overwhelmingly focussed on the Edelstein effect due to the surface states in the presence of a scalar scattering potential. We investigate here the contribution to the SOT due to extrinsic spin-orbit scattering of the surface states, focusing on the case of an out-of-plane magnetization. We show that spin-orbit scattering brings about a sizable renormalization of the field-like SOT, which exceeds 20$\%$ at larger strengths of the extrinsic spin-orbit parameter. The resulting SOT exhibits a maximum as a function of the Fermi energy, magnetization, and extrinsic spin-orbit strength. The field-like SOT decreases with increasing disorder strength while the damping-like SOT is independent of the impurity density. With experimental observation in mind we also determine the role of extrinsic spin-orbit scattering on the anomalous Hall effect. Our results suggest extrinsic spin-orbit scattering is a significant contributor to the surface SOT stemming from the Edelstein effect when the magnetization is out of the plane.
\end{abstract}

\section{Introduction}

The discovery of spin-orbit torques (SOT)~\cite{PhysRevB.66.014407, PhysRevB.80.134403, PhysRevB.86.014416, RevModPhys.91.035004, chernyshov2009evidence,ryu2013chiral, miron2011perpendicular, wadley2016electrical, bandyopadhyay2008introduction,Ramaswamy2018,Shao2021RoadmapOS,Nikolicnikolic2018,zhou2021modulation} has attracted enormous interest from the perspectives of both basic science and industrial applications, as it represents a very promising way to electrically and locally manipulate a magnetic devices at the nanoscale. The strong spin-orbit (SO) interaction ~\cite{RevModPhys.76.323, RevModPhys.87.1213}, which relates electron orbital and spin degrees of freedom, and the exchange interaction, which couples electron spin and a magnetic moment in a ferromagnet, combine to produce a powerful mechanism for manipulating magnetic spin textures ~\cite{dieny2020opportunities, shi2019all, macneill2017control, kent2015new}. The general form of the non-equilibrium spin polarization ${\bf s}({\bf r},t)$ can be deduced by considering moving electrons in the presence of both an inhomogeneous non-equilibrium magnetization ${\bf M}({\bf r},t)$ and a electric field ${\bf E}(t)$. The gradient expansion of ${\bf s}({\bf r},t)$ can be written in terms of ${\bf M}$ and ${\bf E}$ by assuming that the fields are slow and smooth on the spatial and temporal scales set by the electron scattering time and the mean free path. Thus we have $s_{i}=K_{ij}E_j+R_{ij}^{kl}E_j \nabla_l M_k+...$ where the tensor $K$ leads to the spin-orbit torque via e.g. Rashba or Dresselhaus SO interactions~\cite{PhysRevB.95.094401}. Note that without SO interactions, simple symmetry considerations lead to zero $K$. The tensor $R$ represents spin-transfer torques that occur when a spin-polarized current flows from a paramagnet through a perfect interface into a ferromagnet and the direction of the spin-transfer torque depends on the polarity of the current flow \cite{PhysRevB.96.045428,Ralph2008}.  Positive current, electrons flowing from the free layer to the fixed layer, promotes anti-parallel alignment between the two ferromagnetic layers, while negative current, electrons flowing from the fixed layer to the free layer, promotes parallel alignment. The origin of the spin transfer torque is a transfer of spin angular momentum from the conduction electrons to the magnetization of the ferromagnet, and the origin of the angular momentum transfer is absorption of the transverse spin current by the interface ~\cite{PhysRevB.66.014407}. In addition, both tensors strongly depend on the boundary states and impurity potential scattering~\cite{rozhansky2021asymmetric}. For hybrid magnetic/TI systems, large SOTs and magnetization switching have been demonstrated~\cite{mellnik2014spin,fan2014magnetization,PhysRevLett.114.257202,PhysRevLett.119.077702,Fan2014,Wang2017,Li2019,liu2021,fukami2016magnetization,culcer2009,Ramaswamy2019,pan2022efficient}, as well as high-efficiency spin-to-charge signal conversion and spin-pumping~\cite{PhysRevLett.113.196601,PhysRevLett.116.096602}.

Considerable controversy continues to surround the origins of the SOT in topological insulators. The relative contributions of surface and bulk states have been under scrutiny \cite{Ghosh2018}, including a recent study by some of us \cite{cullen_2022}. Our focus in the present work is on the surface states. It is known that these contribute to the spin torque via the Rashba-Edelstein effect~\cite{edelstein1990spin, mihai2010current, mellnik2014spin, Dc2018} and via the spin transfer torque caused by the inhomogeneity in the magnetization~\cite{Kurebayashi2019,siu2018}.
With the exception of the brief discussion in Ref.~\cite{Ghosh2018}, the role of extrinsic spin-orbit scattering in the surface state contribution has not been addressed systematically, a fact that we seek to remedy here. Our study is motivated by the knowledge that, in materials with strong band structure SO interactions such as topological insulators, extrinsic spin-orbit coupling in the impurity scattering potential is necessarily large. Therefore, it is normal to expect a large extrinsic contribution in TIs ~\cite{bonell2020, bhattacharjee2022effects} because SO dominates the surface state spectrum of TI. Extrinsic spin-orbit scattering results in the deflection of electrons via two extensively studied mechanisms. The first is skew scattering~\cite{d1971possibility, PhysRevLett.83.1834, PhysRevLett.95.166605}. The skew scattering mechanism consists of asymmetric scattering of up and down spins. For a scalar band structure it arises in the second Born approximation, though for topological insulators and systems with strong band structure SO interactions in general skew scattering is already present in the first Born approximation. The second mechanism is side jump scattering~\cite{PhysRevB.2.4559}, which describes the lateral displacement of the electron centre of mass during scattering events, again having opposite signs for up and down spins. The surface contribution to the SOT in a realistic topological insulator/ferromagnet structure is determined by the chiral spin textures as well as by scalar and spin-orbit scattering, and their individual roles need to be elucidated.

\begin{figure}
\centering
\includegraphics[scale=0.63]{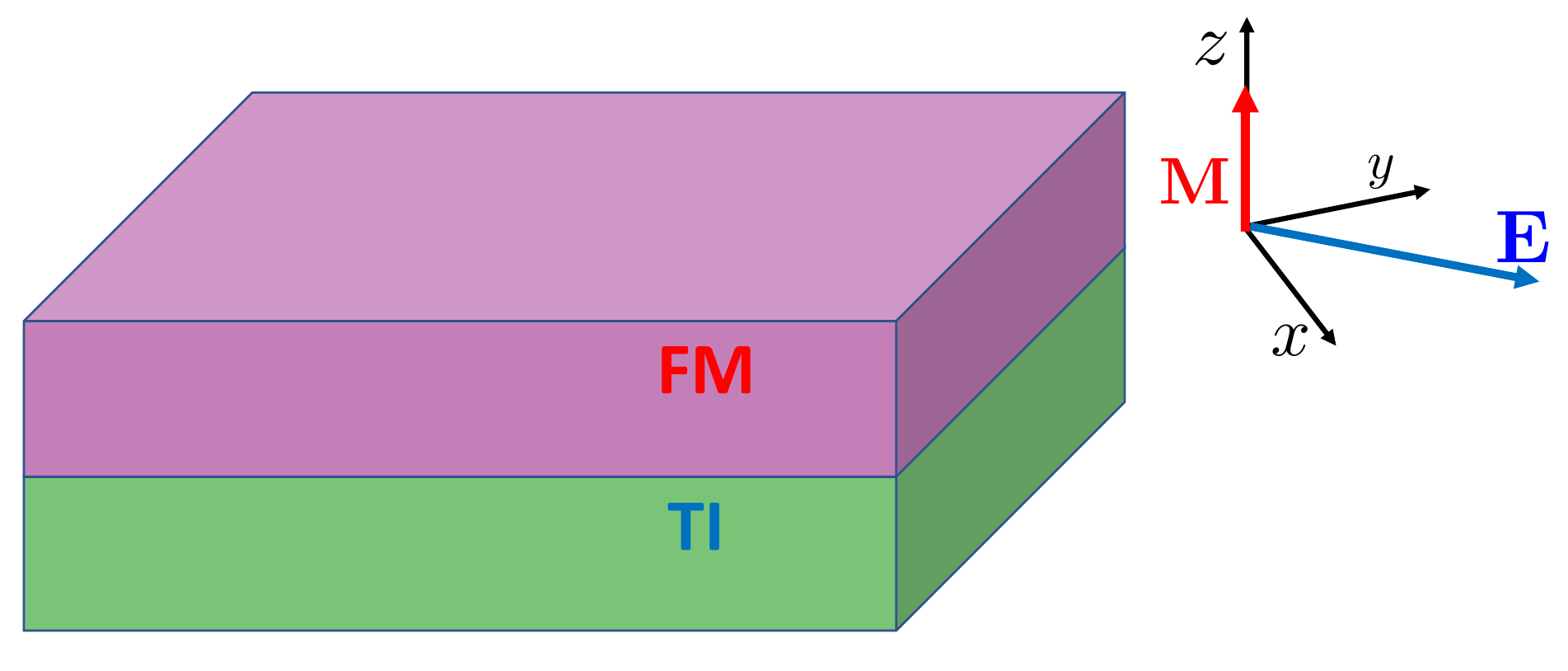}
\caption{Proposed schematic for an FM/TI device with a magnetization perpendicular to the surface of the TI, ${\bf M}=M{\hat z}$. At the interface FM/TI surface $x-y$ plane, there is an applied electric field, {\bf E}.}
\label{device}
\end{figure} 

In this paper, we investigate the SOT in a hybrid system comprising spin-momentum locked massless Dirac fermions in topological insulators ~\cite{RevModPhys.82.3045, RevModPhys.83.1057, moore2010birth, shen2012topological, franz2013topological,Chang2015} interacting with a ferromagnetic layer~\cite{Vobornik2011}, taking into account both intrinsic and extrinsic SO interactions. The magnetization is taken to be out of the plane, as in Fig.~\ref{device}, while the in-plane case will be considered in a future publication. Although the strength of the extrinsic spin-orbit coupling is not known for TI, its form is straightforwardly deduced by symmetry considerations, and estimates are based on the regime of applicability of the theory. We determine the SOT in the presence of a homogeneous magnetization, building on the formalism introduced in Ref. \cite{PhysRevResearch.4.013001} by considering extrinsic impurity scattering in the broader context of ferromagnetic materials. The microscopic SOT is decomposed into two parts: a damping-like SOT  and a field-like SOT around momenta at the interface. Our calculations show that the field-like and damping-like SOTs are dependent on the Fermi energy, the external magnetization strength and the extrinsic impurity scattering and in addition, SOT is strongly nonlinear in terms of the extrinsic SO scattering for not small Fermi energy, given by Eq.~\ref{eq:tau}. At a low extrinsic SO scattering strength, $\lambda$, as well as $M/\varepsilon_F$, the field-like SOT behaves like $\tau _{FL}\sim{\varepsilon _{F} M \tau}(1-2\lambda)$ and the damping-like torque is given by $\tau _{DL} \sim{M^2(1+\lambda)/\varepsilon_{F}}$ where $\tau$ is the electron relaxation time. Moreover, they can be tuned to their maximum value by adjusting the $M$, $\varepsilon _{F}$ and $\lambda$. The field-like SOT component decreases in a disordered system, despite the fact that the strength of damping-like SOT component is independent of the impurity density. 

Since in a realistic sample individual contributions to the SOT are difficult to disentangle, we also consider the fingerprints of extrinsic SO scattering in the anomalous Hall effect (AHE)~\cite{PhysRevLett.93.206602, RevModPhys.82.1539, Liu_AnnRev2016, PhysRevLett.88.207208, Yu_2010,Burkov_2014}. We show that the anomalous Hall resistivity is influenced by extrinsic SOT scattering, and its parameter dependence may offer clues as to the importance of extrinsic processes in a particular sample. Whereas the anomalous Hall conductivity is independent on the impurity in a case of short-range impurity scattering, it depends a highly nonlinear on impurity parameter in the case of the extrinsic SO scattering.  

This paper is organized as follows. In Sec. ~\ref{Intro}, we present the model Hamiltonian in the presence of extrinsic SO scattering. The quantum kinetic theory is discussed in Sec. ~\ref{QKT}, including the electric field correction to the scattering term and corrections to the density matrix up to the sub-leading term in the impurity density. The SOT and AHE in the presence of short-range and extrinsic SO scatterings are presented in Secs.~\ref{SOT} and \ref{AHE}, respectively. Finally, we summarize our main results in the concluding part of the manuscript.

\section{Model Hamiltonian and theory}~\label{Intro}

Topological insulators exhibit strong SO coupling both in the band structure and in the impurity potentials. We consider a 2D topological insulator in the presence of magnetization $M$ where its direction is perpendicular to the plane: a TI is placed on a ferromagnet with magnetization perpendicular to the 2D plane as in Fig.~\ref{device}, and thus the magnetization opens a gap in the surface state spectrum of the TI~\cite{PhysRevLett.104.146802, PhysRevB.81.241410, PhysRevB.96.014408}. 
The exchange coupling between the magnetic moments and the conduction electron spins is represented by the term $M{\sigma_z}$ where ${\bf \sigma}$ is the vector of Pauli matrices representing the spin operators of conduction electrons. The band Hamiltonian that describes low-energy excitations in the surface of 3D topological insulators is given by~\cite{Liu2010,Hasan2010}
\begin{equation}
H_0={\hbar}{v_F}({k_x}{\sigma_y}-{k_y}{\sigma_x})+M{\sigma_z},
\end{equation} 
where $v_F$ is the effective Fermi velocity and ${\bf k}$ is the momentum vector of electrons. The eigenvalues of $H_0$ are $\varepsilon_{\bf{k}}=\pm\sqrt{(\hbar{v_F}k)^2+M^2}$, where $\pm$ labels conduction/valence band and the eigenstates are
$
\langle{u^{s}_{\bf k}}|=\frac{1}{\sqrt{2}}
(e^{i\theta_{\bf k}}\sqrt{1+s\xi_{\bf{k}}}, -s i \sqrt{1-s\xi_{\bf{k}}})
$
where ${\xi_{\bf{k}}}=M/\lambda_{\bf {k}}$ with $\lambda_{\bf{k}}=\sqrt{(\hbar{v_F}k)^2+M^2}$, $s=\pm{1}$,
 and $\theta_{k}=\arctan(k_y/k_x)$.

The total Hamiltonian describing the conduction electrons is
\begin{equation}\label{eq:Eq.1}
H=H_0+V({\bf r})+U(\bf r),
\end{equation}
where $V(\bf r)$ represents the electrostatic potential which has the form $V(\bf r)=e{\bf E}\cdot\bf r$, implying a uniform electric field ${\bf E}$, which corresponds to the overwhelming majority of experimental setups. Also, $U=U(\bf r)$ represents the disorder scattering potential for the conduction electrons.  
In order to capture general behavior of SOT in the system, we consider short-range scattering together with extrinsic SO scattering. For short-range scalar scattering, we consider $U_0({\bf r})=\sum_{i}{u_0}\delta({{\bf r}}-{{\bf R}}_i)$, where ${\bf R}_i$ indicate the random locations of the impurities and $u_0$ is a parameter that measures the strength of the disorder potential~\cite{PhysRevResearch.4.013001}. The average over impurity configurations $\langle U_0({\bf r}) \rangle=0$. Extrinsic SO scattering~\cite{sinitsyn2007semiclassical} is contained in the following term
\begin{eqnarray}
U_{{so}}({\bf r}) = \lambda_0{{\bf \sigma}}\cdot{{\bf \nabla}} U_0({\bf r}) \times{\bf p},
\end{eqnarray}
where ${\bf p}$ is the momentum operator and $\lambda_0$ is the effective extrinsic spin-orbit impurity scattering strength. The matrix elements of the impurity potentials in reciprocal space are  
$U^{ss'}_{{ \bf k}{\bf k}'}=\langle{s,{\bf k}}|U_0({\bf r})|s',{\bf{k}'}\rangle+\langle{s,\bf{k}}|U_{{so}}({\bf r})|{s',{\bf k}'}\rangle.$
The extrinsic SO scattering term can be assumed as a random effective magnetic field, which depends on an electron’s incident and scattered wave vectors. $\lambda_0$ can have either sign. Moreover, it is assumed that $\lambda_0 \ll 1$, and this term is typically treated in perturbation theory. Finally, the disorder potential including scalar and SO impurities reads as
\begin{eqnarray}
U^{ss'}_{\bf{k}\bf{k}'}={u_0}\sum_{i}{e^{i{\bf q}\cdot{\bf R}_i}}\langle{{u}^{s}_{\bf k}}|\Big[\sigma_0+i{\lambda_0}{\bf \sigma}\cdot({\bf k}\times{\bf k'})\Big]|{{u}^{s'}_{\bf k'}}\rangle,
\end{eqnarray}
where $\sigma_{0}$ is the 2$\times$2 in spin space and ${\bf q}={\bf k'}-{\bf k}$. If $\vert{\bf  k}\vert=\vert{\bf  k'}\vert=k_{F}$ and assuming that there are no spatial correlations in the disorder potential, the scattering matrix elements will be 
\begin{eqnarray}
U^{ss'}_{\bf{k}\bf{k}'}={u_0}\langle{{u}^{s}_{\bf k}}|\Big[\sigma_0+i{\lambda}{\sigma}_z{\sin{\gamma}}\Big]|{{u}^{s'}_{\bf k'}}\rangle,
\end{eqnarray}
where $\lambda=\lambda_{0}{k_{F}^2}$ and $\gamma = \theta_{\bf k'}-\theta_{\bf k}$. Averaging over impurity configurations, the first-order term in the potential vanishes due to the randomness of the impurity locations as usual, while the second-order term
\begin{eqnarray}
\langle{U}^{mm'}_{{\bf k{\bf k'}}}{U}^{m''{m'''}}_{{\bf k'{\bf k}}}\rangle={n_i}{U}^{mm'}_{{\bf k{\bf k'}}}{U}^{m''m'''}_{{\bf k'{\bf k}}},
\end{eqnarray}
where $m(m')$ and $m''(m''')$ are spin indices and $n_i$ indicating the impurity concentration. 
\begin{figure}
\centering
\includegraphics[scale=0.73]{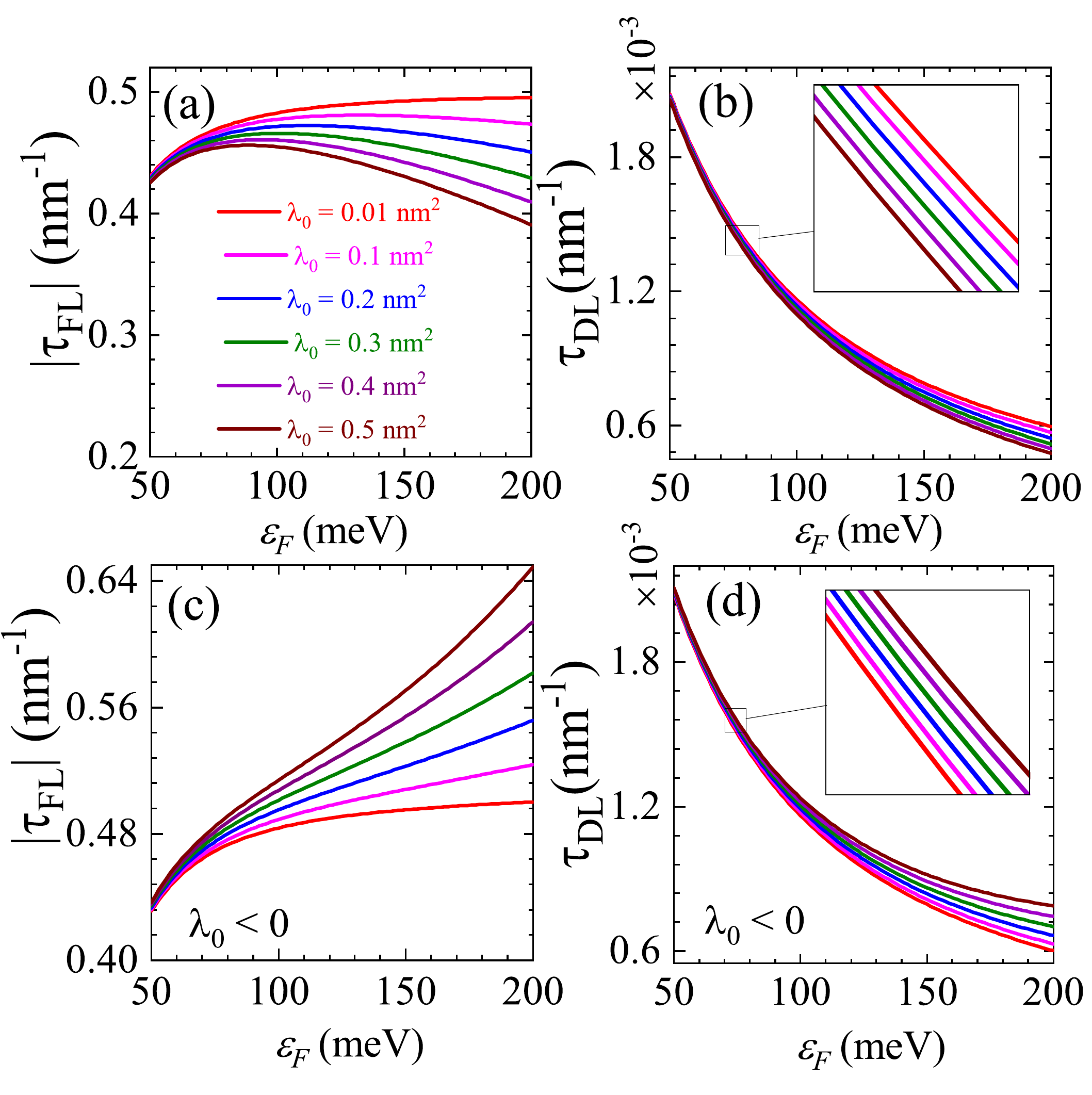}
\caption{Field-like SOT as a function of the Fermi energy for the various (a) positive (c) negative values of the extrinsic SO scattering strength. Damping-like SOT as a function of the Fermi energy for the various (b) positive (d) negative values of the extrinsic SO scattering strength. The exchange field due to magnetization is considered as $M=$10 meV.}
\label{fig.3}
\end{figure}


\section{Quantum Kinetic Theory}~\label{QKT}

The density matrix $\rho=f_{\bf k}+g_0$ characterizes the system considered in Eq. (\ref{eq:Eq.1}), where $f_{\bf k}$ is averaged over disorder configurations and has matrix elements connecting different bands, and $g_0$ is the fluctuating part.
For the disorder-averaged component in equilibrium, the density matrix obeys the quantum kinetic equation \cite{vasko2006quantum}
\begin{equation}\label{eq:QKE}
    \frac{\partial f_{\bf k}}{\partial t} + \frac{i}{\hbar} \, [H_{0{\bf k}}, f_{\bf k}] + J_{0}(f_{\bf k}) = \frac{e{\bf E}}{\hbar}\cdot\frac{Df_{\bf k}}{D{\bf k}} - J_E(f_{\bf k}),
\end{equation}
where the covariant derivative is as follows: $\frac{Df_{\bf k}}{D{\bf k}} = \frac{\partial f_{\bf k}}{\partial{\bf k}} -i[{\bf \mathcal{R}}_{\bf k}, f_{\bf k}]$ where ${\bf \mathcal {R}} ^{ss}_{\bf k}=i\langle {{u}^{s}_{\bf k}}|\nabla_{\bf k}{{u}^{s}_ {\bf k }}\rangle$ is the Berry connection. In order to solve Eq.~(\ref{eq:QKE}), the density matrix is divided into band-diagonal and off-diagonal parts, namely $f_{\bf k} = n_{{\bf k}} + S_{{\bf k}}$~\cite{Culcer_2017, Sekine_2017}. The band diagonal term $n_{{\bf k}}$ represents the fraction of carriers in a given band and is essentially the solution to the ordinary Boltzmann equation, while $S_{{\bf k}}$ contains the effect of interband coherence. $J_{0}(f_{\bf k})$ is the collision integral due to the impurity potential, which is defined as
$J_0(f_{\bf k})=\frac{i}{\hbar}\langle [U,g_0]\rangle,$
where $g_0$ in terms of Green's functions can be expressed as
\begin{equation}\label{eq:Eq.23}
g_0=\frac{1}{2\pi i}\int^{\infty}_{0}d\epsilon [G^{R}_{0}(\epsilon)U G^{A}_{0}(\epsilon),f_{\bf k}]. 
\end{equation}
Also, $G^R_0(\epsilon)$ is a free retarded Green function obtained as
\begin{equation}
G^R_0(\epsilon)=-\frac{i}{\hbar}\int^{\infty}_{0} dt e^{-iH_0t/\hbar} e^{i\epsilon t/\hbar} e^{-\eta t},
\end{equation}
where we introduce the factor $e^{-\eta t}$ to ensure convergence and the advanced Green function is defined as $G^{A}_{0}(\epsilon)=G_{0}^{R,\dagger}(\epsilon)$. 

Now we want to include the effect of the driving electrostatic potential up to linear order~\cite{PhysRevResearch.4.013001}. Adding an electric field to the Hamiltonian implies a correction of the function $g$. Moreover, $J_E(f_{\bf k})$ is a correction of the collision integral due to the external electric field and is given by 
$J_E(f_{\bf k}) = \frac{i}{\hbar}\langle [U,g_E] \rangle,$ 
where the $g_E$, which reflects it is the electric field correction and off diagonal in the momentum and in the band index, defined as 
\begin{equation}
\label{Eq:FrequencyDomain_g(E)}
g_E  = \frac{1}{2\pi i}\int^{\infty}_{0}d\epsilon G^{R}_{0}(\epsilon)[V, g_0 ] G^{A}_{0}(\epsilon).
\end{equation}
To perform a linear response, one can use Eq.~(\ref{eq:Eq.23}) as a function of the equilibrium distribution function $f_0(\epsilon)$.
If we expand the density matrix $f_{\bf k}$ to linear order in the electric field as $f_{\bf k}^{mn} = f_0(\epsilon ^m_{\bf k})\delta_{ mn} + n_{E}(\epsilon^{m}_{\bf k})+S_{E{\bf k}}^{mn}$, the kinetic equation (i.e., Eq.~(\ref{eq:QKE})) for the diagonal part of the density matrix is written as
\begin{equation}
[J_{0}(n_E)]^{mm}_{{\bf k}} = \frac{e\bf E}{\hbar}\cdot \frac{\partial f_{0}(\epsilon^{m}_{\bf k})}{\partial \bf k}. 
\end{equation}
Notice that the kinetic equation is considered to linear order of the electrical field and basically $n_{E}$ begins at order $-1$ in terms of small parameter $\hbar/\varepsilon_F \tau$~\cite{Culcer_2017} where $\tau$ is the momentum relaxation time. If the band dispersion is isotropic (e.g., band dispersion for topological insulators), one can assume $[J_{0}(n_E)]^{mm}_{\bf k}=n_{E}(\epsilon^{m}_{\bf k})/\tau_{tr}(\bf k)$, so the equation above becomes as
\begin{equation}
\label{eq:nDrude}
n_{E}(\epsilon^{m}_{\bf k}) =\tau_{tr}({\bf k}) \frac{e\bf E}{\hbar}\cdot 
\frac{\partial \epsilon^{m}_{\bf k}}{\partial \bf k} 
\frac{\partial f_{0}(\epsilon^{m}_{\bf k})}{\partial \epsilon^{m}_{\bf k}}.
\end{equation}
Using the expression of $J_0(f_{\bf k})$, the diagonal part of the Born approximation collision integral can also be written as 
\begin{equation}
\label{eq:transportTime}
[J_{0}(n_E)]^{mm}_{{\bf k}} = \frac{2\pi}{\hbar}\sum_{m',\bf k'}
\langle U^{mm'}_{\bf k \bf k'} U^{m'm}_{\bf k' \bf k}\rangle
\left[n^{mm}_{E{\bf k}}-n^{m'm'}_{E{\bf k'}}\right]\delta(\epsilon^{m}_{\bf k}-\epsilon^{m'}_{\bf k'}).
\end{equation}
According to Eq.~(\ref{eq:nDrude}) and ${\bf v}=({1/\hbar}){\partial \epsilon^{m}_{\bf k}}/{\partial\bf k}={\bf k}/m$, we can calculate $n_{E}(\epsilon^{m}_{\bf k})$ in the electric field.

We assume transport time is angular-independent or ${\bf k}$ direction independent. Substituting the above equation in Eq.~(\ref{eq:transportTime}) and assuming an elastic scattering $\varepsilon_{\bf k'}=\varepsilon_{\bf k}$ when the direction of electrical field is $\hat{k}$, the transport time obtains as
\begin{equation}\label{eq:Eq.33}
 \frac{1}{\tau_{tr}(k)}=\frac{2\pi}{\hbar}\sum_{m',\bf k'}
\langle U^{mm'}_{\bf k \bf k'} U^{m'm}_{\bf k' \bf k}\rangle
\left[1-\cos(\theta_{\bf k}-\theta_{\bf k'})\right]\delta(\epsilon^{m}_{\bf k}-\epsilon^{m'}_{\bf k'}).
\end{equation}
As discussed in \cite{Culcer_2017}, $S^{mm'}_{E \bf k}$ starts as order $0$ based on an expansion in the small parameter $\hbar/(\varepsilon_F\tau)$. The kinetic equation for the off-diagonal density matrix is written as
\begin{equation}\label{QKE}
\frac{\partial S^{(0)}_{E \bf k}}{\partial t} + \frac{i}{\hbar} \, [H_{0{\bf k}}, S^{(0)}_{E \bf k}]= D_{\bf E} + D^{'}_{\bf E},
\end{equation}
where $D$ is the intrinsic driving term and $D'$ is the anomalous driving term, containing disorder renormalizations of the response which are in part equivalent to vertex corrections.
Having solved the above equation,
the intrinsic and extrinsic contributions are defined as
\begin{eqnarray}\label{eq:Eq.39}
&&[S^{(0)}_{\bf E}]_{\bf k, int}^{mm'}={\frac{\hbar D^{mm'}_{\bf E \bf k}}{i(\epsilon_{\bf k}^{m}-\epsilon_{\bf k}^{m'}-i\eta)}}, \\
&&[S^{(0)}_{\bf E}]_{\bf k, ext}^{mm'}={\frac{\hbar D'^{mm'}_{\bf E \bf k}}{i(\epsilon_{\bf k}^{m}-\epsilon_{\bf k}^{m'}-i\eta)}}.
\end{eqnarray}

The diagonal subleading distribution function $n^{(0)}_{E}$ can be calculated as
$J (n^{(0)}_{\bf E})= -J_0 (S^{(0)}_{\bf E}) - J_{\bf E} (f_0 (\bf E))$
where the right hand side plays as the driving term. By solving this equation, we get two different contributions to the subleading diagonal density matrix as $n^{(0)}_{\bf E}=n^{(sj)}_{\bf E}+n^{(sk)}_{\bf E}$ and one can write
$J (n^{(sj)}_{\bf E})= - J_{\bf E} (f_0 (\bf E))$, 
and $J (n^{(sk)}_{\bf E})= - J_{0}(S^{(0)}_{\bf E})$
with $n^{(sj)}_{\bf E}$ and $n^{(sk)}_{\bf E}$ are due to the side jump and skew scattering contributions, respectively. We can also decompose the off-diagonal density matrix into $S^{mn}_{\bf E}=S^{mn}_{int}+S^{mn}_{ext}$.
Having calculated the scattering terms and after straightforward but lengthy calculations, we find
\begin{eqnarray}\label{eq:Eq.92}
n^{(sk)}_{E}(\epsilon^{m}_{\bf k})=eZ^2_{\bf k}(1-\xi^2_{\bf k}){\bf E}\cdot\hat{\bf \theta} {\sigma_z},
\end{eqnarray}
and,
\begin{eqnarray}\label{eq:Eq.141}
n^{(sj)}_{E{\bf k}}=Z_{\bf k}\left[{1+0.5\lambda(1+\xi^2_{\bf k})(1-\xi^2_{\bf k})^{-1}}\right]{\bf E}\cdot\hat{\bf k}
\end{eqnarray}
where 
\begin{equation}
Z^{\alpha}_{\bf k}=\frac{8\hbar{v_F}}{\lambda_{\bf k}}\frac{\xi_{\bf k}(1-\xi^2_{\bf k})^{\frac{1}{2}}\delta(\epsilon_{F}-\epsilon^{+}_{\bf k})}{\chi^{\alpha}(\lambda)},
\end{equation}
here $\chi(\lambda)=4(1+3\xi^2_{F})+{\lambda}^2(5+3\xi^{2}_{F})+8\lambda(1-\xi^2_{F})$. Furthermore, the extrinsic and intrinsic contributions of the off diagonal term is given by
\begin{equation}\fl
S_{ext}=\frac{\hbar}{2\lambda_{\bf k}}\frac{e\tau_{tr}}{4{\tau}}{v_F}(1-\xi^2_{\bf k}){\delta(\epsilon_{F}-\epsilon^{+}_{\bf k})}
\left[(3-2\lambda+\frac{3}{4}{\lambda}^2){\xi_{k}}{\bf E}\cdot{ \hat{\bf \theta}}{\sigma_{y}}-(1-\frac{3}{4}{\lambda}^2){\bf E}\cdot{\hat{\bf k}}{\sigma_{x}}\right],
\end{equation}
and
\begin{eqnarray}\label{eq:int}
S_{int}=\frac{e\hbar{v_F}}{(2\lambda_{\bf k})^2}[f_0(\epsilon^{+}_{\bf k})-f_0(\epsilon^{-}_{\bf k})]
\times\left({\bf E}\cdot{\hat{\bf k}}{\sigma_{x}}-{\xi_{k}}{\bf E}\cdot{ \hat{\bf \theta}}{\sigma_{y}}\right),
\end{eqnarray}
where $\tau_{tr}={\tau}/\chi(\lambda)$ with $1/\tau={{n_i}\pi{u^2_0}\rho(\epsilon_{\bf k})}/\hbar$, and $\rho(\epsilon_{F})=\epsilon_{F}/2\pi{\hbar^2}{v^2_F}$ is the density of
states. Here, $\hat{\bf k}=\hat{x}\cos{\theta}+\hat{y}\sin{\theta}$ and $\hat{\bf \theta}=-\hat{x}\sin{\theta}+\hat{y}\cos{\theta}$. 

It is clear that the extrinsic SO scattering renormalizes physical quantities given by Eqs. (\ref{eq:Eq.92}-\ref{eq:int}) through the parameter $\lambda$.

 \section{SPIN DENSITY AND SPIN ORBIT TORQUE}\label{SOT}

The spin density and spin-orbit torques in topological insulators with out-of-plane magnetization are determined in this section. The spin density has five contributions: a leading contribution at the Fermi surface analogous to the Drude conductivity, an intrinsic contribution that takes into account the Fermi sea, an extrinsic contribution due to the Fermi surface that is formally zeroth order in the impurity potential strength, a side-jump contribution at the Fermi energy which includes the electric field correction to the collision integral, and a skew scattering contribution. We determine each of these in turn below. 

Let us calculate the leading order contribution to the spin density. It is
\begin{eqnarray}
\langle{s_x}\rangle=\sum_{m, \bf k}{s^{mm}_{{\bf k}, x}}{n(\epsilon^{m}_{\bf k})}
\end{eqnarray}
With the leading order density matrix ${n(\epsilon^{m}_{\bf k})}=-e{\tau_{tr}}{\bf E}\cdot{v^{m}_{\bf k}}\delta(\epsilon_{\bf k}^{m}-\epsilon_{F})$
, we find the spin density as
\begin{eqnarray}
\langle{s_x}\rangle^{{leading}}=\frac{1}{2}(1-\xi^2_{k_{F}}){e\tau_{tr}}{E_{y}}{v_F}{\rho(\epsilon_{F})}.
\end{eqnarray}
Note that the contributions including $v_{x}$ due to $\int^{2\pi}_{0}{d\theta_{\bf k}}{\sin{\theta}_{\bf k}}{\cos{\theta}_{\bf k}}=0$ is zero. In a similar manner we find
\begin{eqnarray}
\langle{s_y}\rangle^{{leading}}=-\frac{4e\tau{E_x}{v_F}\rho(\epsilon_{F})(1-\xi^{2}_{\bf k_F})}{\chi(\lambda)}.
\end{eqnarray}

Let us now consider the off-diagonal intrinsic density matrix contribution. First, the off-diagonal spin expectation value is 

\begin{eqnarray}
\sum_{\bf k}2{Re}[{s^{-+}_{x}}S_{int}^{+-}]=\frac{e{E_x}\hbar{v_F}M}{2}\frac{\rho(\epsilon_{F})}{\epsilon^{2}_{F}}.
\end{eqnarray}
Furthermore, we consider the off-diagonal extrinsic density matrix contribution as
\begin{eqnarray}
\sum_{\bf k}{{Re}[s^{-+}_{x}S^{+-}_{{ext}}]}={e\hbar{v_F}}\frac{M}{\epsilon^{2}_{F}}{\rho(\epsilon_{F})}{E_x}\frac{(1-\xi^2_{F})(2-\lambda)}{\chi(\lambda)}
\end{eqnarray}
Moreover, the diagonal subleading density function of skew scattering will be as
\begin{eqnarray}
\langle{s_x}\rangle^{(0)}=8e\hbar{v_F}\frac{M}{\epsilon^2_{F}}{\rho(\epsilon_{F})}{E_x}\frac{(1-\xi^2_{F})^2}{\chi^2}
\end{eqnarray}
In oder to find the spin density related to the subleading of side jump, 
\begin{eqnarray}
\langle{{\bf s}}\rangle^{{sj}}=e\hbar{v_F}\rho(\epsilon_{F})\frac{M}{\epsilon^{2}_{F}}\frac{4(1-\xi^{2}_{F})+2\lambda(1+\xi^{2}_{F})}{\chi(\lambda)}{{\bf E}}
\end{eqnarray}
The $z-$component of spin expectation, on the other hand, are found as
$s^{\pm\pm}_{z}=\langle{u^{\pm}_{\bf k}}|{\sigma_{z}}|{u^{\pm}_{\bf k}}\rangle=\pm{\xi_{\bf k}}$ and $ s^{\pm\mp}_{z}=\langle{u^{\pm}_{\bf k}}|{\sigma_{z}}|{u^{\mp}_{\bf k}}\rangle=(1-\xi^2_{\bf k})^{1/2}$ which are angle independent and therefore, the $z$-component spin density vanishes: 
\begin{eqnarray}
&&\langle{s_z}\rangle^{{leading}}=\sum_{\bf k}s^{++}_{{\bf k},{z}}{n(\epsilon^{+}_{\bf k})}+\sum_{\bf k}s^{--}_{{\bf k},{z}}{n(\epsilon^{-}_{\bf k})}\\
&=&-2\sum_{\bf k}{e}{\tau_{tr}}{v_F}{\xi_{\bf k}}(1-\xi^2_{\bf k})^{1/2}\Big({E_x}\sin{\theta_{k}}+{E_y}\cos{\theta_{k}}\Big)\left[\delta(\epsilon^{-}_{\bf k}-\epsilon_{F})+\delta(\epsilon^{+}_{\bf k}-\epsilon_{F})\right]=0.\nonumber
\end{eqnarray}

Now we add all these contributions and get
\begin{eqnarray}\fl
{\langle}{\bf s}{\rangle}={Tr}\{{\bf s}{\langle\rho\rangle}\}=-4e\tau{v_F}\rho(\epsilon_{F})\frac{(1-\xi^{2}_{F})}{\chi(\lambda)}\hat{\bf z}\times{\bf E}
+e{\hbar}{v_F}\rho(\epsilon_{F})\frac{M}{\epsilon^{2}_{F}}\frac{\eta(\lambda)}{2{\chi}^2(\lambda)}\bf{E},
\end{eqnarray}
 where $\eta(\lambda)=\chi^2+8\lambda \xi^{2}_{F}\chi+16(1-\xi^{2}_{F})(1+\chi-\xi_{F}^2)$.
It is worthwhile mentioning that the effect of the SO impurity scattering is explicitly appeared in the results through parameter $\lambda$.

 The spin orbit torque is defined as $\frac{d{\bf m}}{dt}=(2M/\hbar){\bf m}\times\langle{s}\rangle$ where ${\bf m}$ is a unit vector along the ${\hat z}$ direction. With the spin density we already calculated we get 
 \begin{eqnarray}\label{eq:tau}\fl
\frac{d{\bf m}}{dt}=-4{M}e\tau{v_F}\rho(\epsilon_{F})\frac{(1-\xi^{2}_{F})}{\chi(\lambda)}{\bf m}\times(\hat{\bf z}\times{\bf E})
+\frac{e{\hbar}{v_F}\rho(\epsilon_{F})M^2 \eta(\lambda)}{2{\chi}^2}{\bf m}\times\bf{E},
\end{eqnarray}
where we have multiplied by a factor of ${\hbar}/{2}$ the spin density since we traced only the Pauli matrix.   
\begin{figure}
\centering
\includegraphics[scale=0.65]{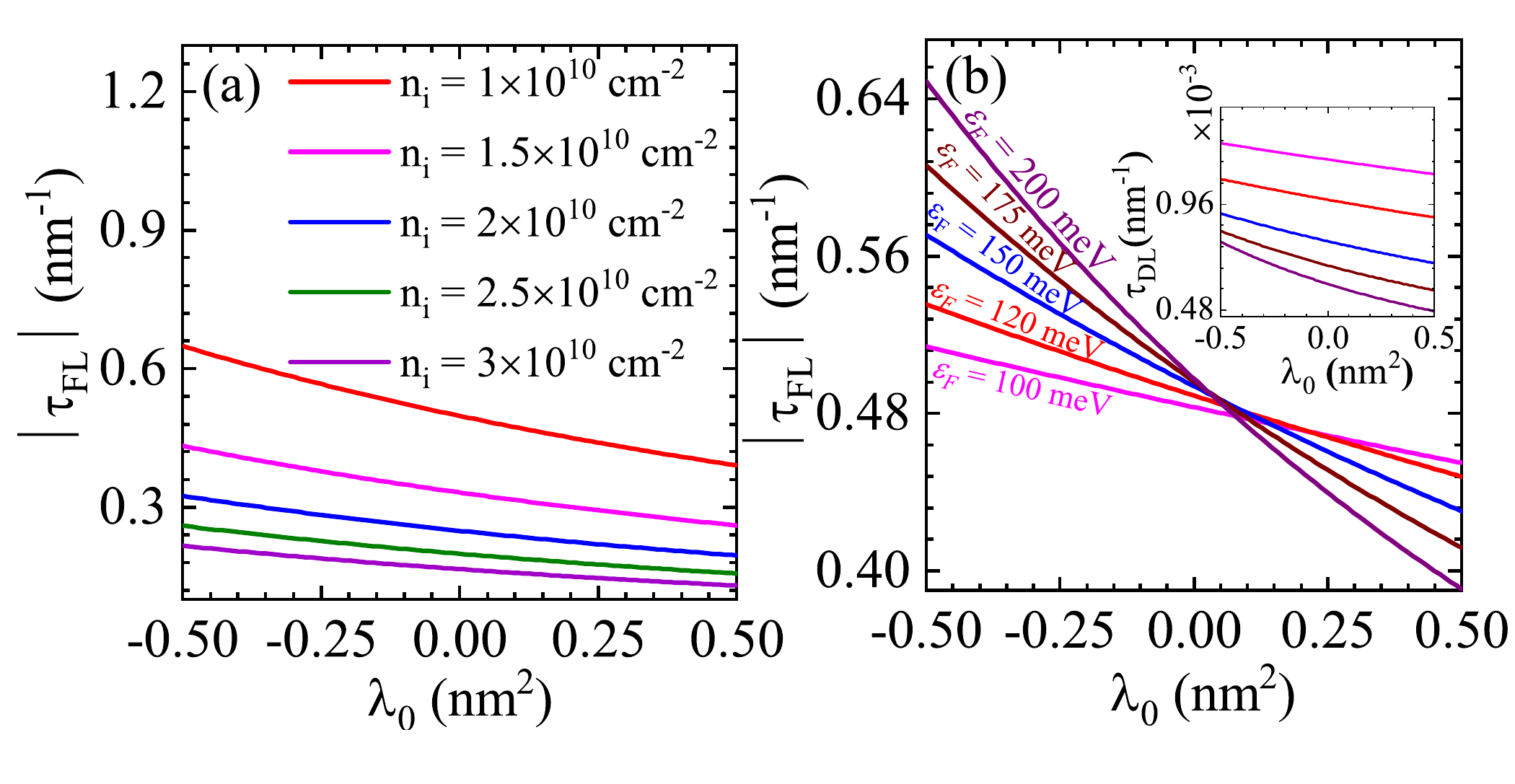}
\caption{Field-like SOT as a function of the extrinsic SO scattering strength, $\lambda_0$ (a) for some distinctive values of impurity density and (b) for different values of the Fermi energy. Here, the damping-like SOT is independent of impurity density. The inset of (b) shows the corresponding damping like SOT. The exchange field due to magnetization is considered as $M=10$ meV.}
\label{fig.4}
\end{figure}

The effective Dirac SOT is defined as $\frac{d{\bf m}}{dt}={\tau}_{FL}{\bf m}\times({\hat {\bf z}}\times{e{\bf E}})+{\tau}_{DL}{m_z}{\bf m}\times{e{\bf E}}$,  
where the first term which is proportional to ${\bf m}\times(\hat{\bf z}\times{e{\bf E}})$ is odd upon magnetization reversal, proportional to the $\tau$ and acts like a field-like torque, while the second term, ${{\bf m}\times{e\bf{E}}}$, is even in magnetization reversal acts like a damping torque. The Rashba-Edelstein effects due to spin-momentum locking on the surface of the topological insulator are responsible for the field-like contribution~\cite{PhysRevB.84.155440}, while the electromagnetic coupling is responsible for the damping-like contribution~\cite{PhysRevB.96.014408}.  According to Eq. (\ref{eq:tau}), we expect ${\tau}_{FL}\gg{\tau}_{DL}$ because the calculations is performed in the regime $\epsilon_{F}\tau/\hbar\gg{1}$. 

Expanding the SOTs for $\lambda\ll{1}$ and skipping terms including $\xi^n_{F}$ for $n>2$, we have
\begin{eqnarray} 
{\tau}_{DL}&\approx&\frac{3M^2}{2\pi\hbar{v_F}\varepsilon_{F}}\Bigg[\Bigg(1+\lambda+\frac{41}{12}\lambda^2\Bigg) - \Bigg(\frac{M}{\epsilon_{F}}\Bigg)^2(4+11{\lambda}+55\lambda^2)\Bigg]\nonumber\\
{\tau}_{FL}&{\approx}&\frac{-{M}\varepsilon_{F}\tau}{2\pi\hbar^2{v_F}}\Bigg[\Bigg(1-2\lambda+\frac{71}{4}\lambda^2\Bigg) - \Bigg(\frac{M}{\epsilon_{F}}\Bigg)^2(4-16\lambda+190\lambda^2)\Bigg]
\end{eqnarray}

We consider a topological insulator (i.e. Bi$_2$Se$_3$) whose Fermi velocity of its massless Dirac fermions is $v_F\simeq{5\times10^5}$m s$^{-1}$. For all figures the impurity density and the amplitude of the scattering potential are given as $n_i=1\times10^{10}$cm$^{-2}$ and $U_0=5 $eVnm$^2$ or otherwise specified.
Figure \ref{fig.3} shows the field-like and damping-like SOTs in terms of Fermi energy for some positive and negative different values of extrinsic SO scattering strength. As seen, the field-like SOT in Figs. \ref{fig.3}(a,c) is larger than the damping-like SOT shown in in Figs. \ref{fig.3}(b,d). As the Fermi energy increases the extrinsic mechanisms including skew and side-jump scatterings become stronger and lead to an increase in the field-like SOT. The extrinsic SO scattering strength plays a crucial role in renormalizing the SOTs so that for positive (negative) $\lambda_0$ will have a decrease (an increase) in the SOTs. In principle, the field-like SOT depends on the density of impurities through the relaxation time. 

\begin{figure}
\centering
\includegraphics[scale=0.82]{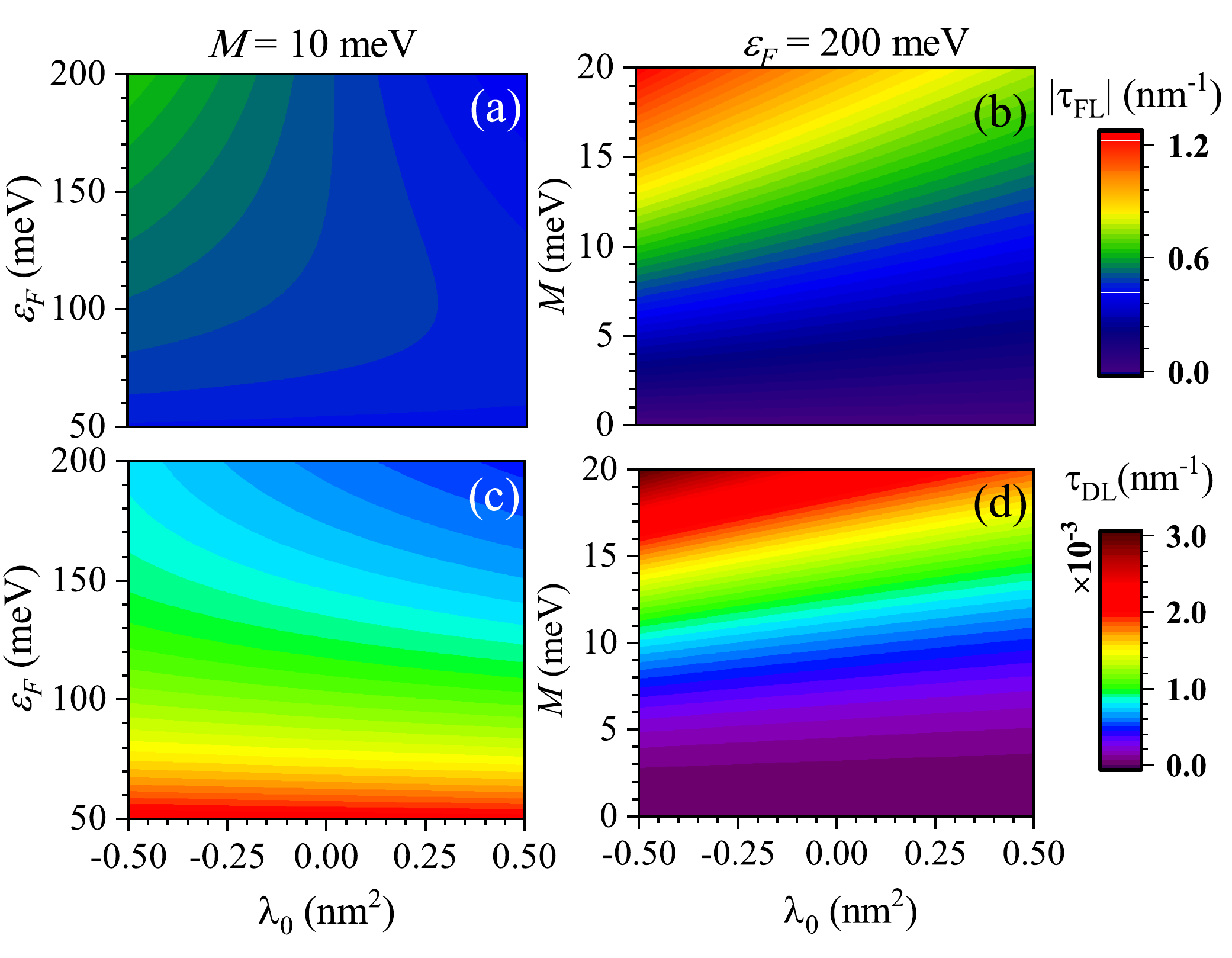}
\caption{Contour plot of the field-like SOT as a function of the extrinsic SO scattering strength and (a) Fermi energy for a constant value of the exchange field strength $M=10$ meV and (b) the exchange field strength for a fixed value of the Fermi energy $\varepsilon_{F}=0.2$ eV. (c), (d) The damping-like SOT is the same as in (a) and (b).}
\label{fig.5}
\end{figure} 

As seen in Fig.~\ref{fig.4}(a), with increasing impurity density, the relaxation time $\tau$ decreases, causing a decrease in the field-like SOT. Notice that the damping-like SOT is independent of impurity density. Moreover, we plotted the damping-like and field-like SOTs as a function of extrinsic SO scattering strength for the distinctive values of Fermi energy in Fig.~\ref{fig.4}(b). It is seen that at larger negative values of $\lambda_0$ and of the Fermi energy the field-like SOT reaches a maximum value. However, the damping-like SOT shows a peculiar behavior as a function of the Fermi energy, decreasing at larger values of $\epsilon_F$. For $\epsilon_{F}=M=10$ meV (i.e., $\xi_{F}=1$) only the intrinsic mechanism contributes to SOTs. When the extrinsic SO scattering strength ($\lambda_0<0$ ) increases other mechanisms such as skew and side-jump scattering contribute to SOTs and lead to an increase in SOTs.

The simultaneous effect of the exchange field strength or Fermi energy and the extrinsic SO scattering strength on the damping-like and field-like SOTs are investigated in the density plots of Fig.~\ref{fig.5}.
The value of the SOT depends strongly on the Fermi energy and magnetization. The field-like SOT increases by with increasing $\varepsilon_F$ or $M$, while the damping-like SOT decreases with increasing Fermi energy. Moreover, as shown in Fig. \ref{fig.5}(a), the field-like SOT behaves nonlinearly in terms of $\lambda_0 \le 0$ for $\varepsilon_F \le 100$ meV at $M=10$ meV, while the damping-like SOT is highly nonlinear for $\lambda_0 \ge 0$ as shown in Fig. \ref{fig.5}(c). Therefore, the SOT behaves highly nonlinearly for large Fermi energy in terms of $\lambda_0$. In addition, the SOT increases with increasing exchange field strength, as shown in Figs. \ref{fig.5}(b) and (d).

\section{Anomalous Hall effect in the presence of extrinsic spin-orbit scattering}
~\label{AHE}

The anomalous Hall conductivity contains four contributions: an intrinsic contribution from the Fermi sea, a contribution due to the extrinsic velocity at the Fermi surface which is nominally of zeroth order in the disorder strength, a side-jump contribution at the Fermi energy including the electric field correction to the collision integral and a contribution due to skew scattering. First we have to determine the expected velocity value as an operator trace Tr($\dot{\bf r}f$), where $\dot{\bf r}=(i/\hbar)[H, {\bf r}]$ represents the Matrix elements of the velocity operator. The Hamiltonian of the system, $V({\bf r})$ commutes with the position operator while we ignore the contribution of the extrinsic SO scattering, which $U({\bf r})$ contributes to the velocity operator. 

The extrinsic velocity in the conduction band is
obtained as
\begin{eqnarray}
\beta^{+}_{\bf k}=\frac{1}{\tau}{\xi_{\bf k}}(1-\xi^2_{\bf k})^{1/2}\frac{\hbar{v_F}}{\lambda_{\bf k}}(1-\lambda){\sigma_{0}}\hat{\bf\theta},
\end{eqnarray}
where $\hat{\bf\theta}=(-\sin{\theta},\cos{\theta})$. Note that the extrinsic velocity is proportional to the impurity density and in contrast to the group velocity, which is a velocity between collisions, the extrinsic velocity includes the effect of disorder on carrier dynamics, and can be read as an effective velocity of the electron after numerous collisions.   

The contribution to the anomalous Hall conductivity related to the skew scattering and side jump corrections is
\begin{eqnarray}\label{eq:Eq.177}
\sigma_{yx}=\sigma^{0}_{yx}\Bigg[\frac{\eta(\lambda)+4\lambda(1-\xi^{2}_{F})\chi(\lambda)}{4\chi^2(\lambda)}\Bigg], 
\end{eqnarray}
where $\sigma^{0}_{yx}=\frac{e^2}{\pi\hbar}\frac{M}{\epsilon_{F}}$. While the Hall conductivity is independent on the impurity in a case of short-range impurity scattering, it depends a highly nonlinear on $\lambda$ in the case of the extrinsic SO scattering. The Hall conductivity can be expand for $\lambda\ll{1}$ up to $\lambda^2$ and $\xi_{F}<1$ (i.e., $\epsilon_{F}>M$), and we thus obtain
\begin{eqnarray} 
\sigma_{yx}\approx\frac{\sigma^{0}_{yx}}{4}\Bigg[\frac{6}{1+6\xi^2_{F}}-\frac{11}{1+9\xi^2_{F}}\lambda-\frac{1.5\xi_{F}+4}{1+12\xi^2_{F}}\lambda^2\Bigg].
\end{eqnarray}
It can be found from above equation, the Hall conductivity decreases (increases) for $\lambda_{0}>0(\lambda_{0}<0)$ as $\sigma_{yx}\sim{\epsilon_{F}M}(\epsilon^2_{F}-M^2)\lambda_{0}$. The longitudinal conductivity
\begin{eqnarray}\label{eq:Eq.181}
\sigma_{xx}=\frac{2e^2\epsilon_{F}\tau}{\pi\hbar^2}\frac{(1-\xi^{2}_{F})}{\chi(\lambda)}.
 \end{eqnarray}
Note that our calculations were performed in the $\epsilon_{F}\tau/\hbar\gg{1}$ regime, i.e. as expected from the $\sigma_{xx}\gg\sigma_{xy}$ equations above. The result for the longitudinal conductivity agrees with the literature $\sigma_{xx}\sim(e^2/h)(v_{F}\tau_{tr}/2)$ (see Eq. (12) in Ref. \cite{liu2017weak}) obtained from the Keldysh-Green's function. 
\begin{figure}
\centering
\includegraphics[scale=0.74]{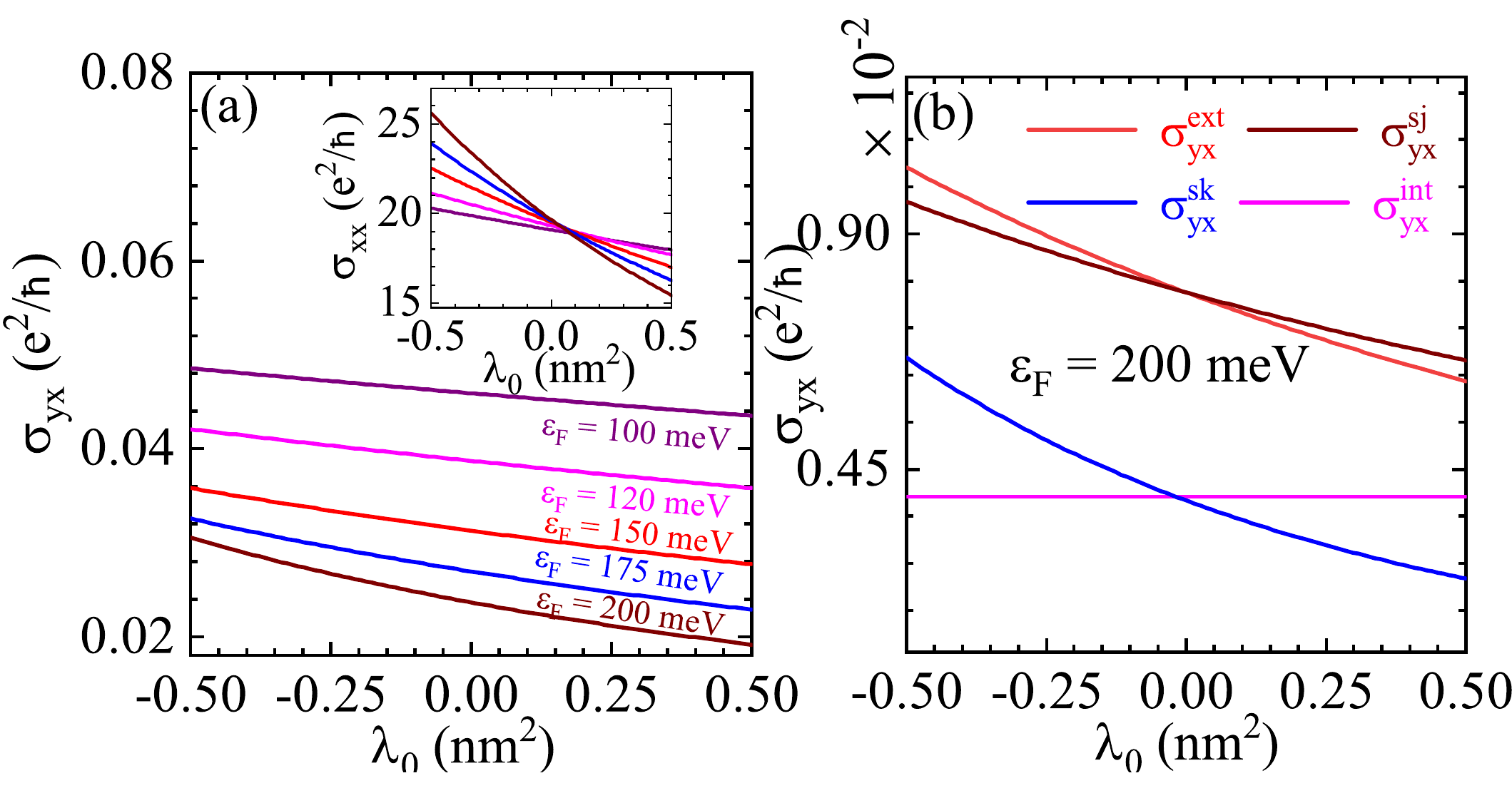}
\caption{(a) The anomalous Hall conductivity in term of $\lambda_{0}$ for different values of the Fermi energy. The inset of (a) shows longitudinal conductivity of a topological insulator. (b) Different contributions to the Hall conductivity for a fixed value of the Fermi energy $\varepsilon_{F}=200$ meV. Notice that the exchange field due to magnetization is $M=10$ meV.}
\label{fig.1}
\end{figure} 

The anomalous Hall conductivity as a function of the extrinsic SO scattering strength $\lambda_{0}$ (which is defined in $\lambda=\lambda_{0}{k^2_F}$) for the different values of the Fermi energy is shown in Fig.~\ref{fig.1}.  The anomalous Hall conductivity is independent of the extrinsic SO scattering strength at $\epsilon_{F}=M=10$ meV (i.e. $\xi_{F}=1 $) since it is only determined by the intrinsic contribution (i.e. $\sigma_{yx}=\sigma ^{{int}}_{yx}$). In this case,  the Berry curvature arises from electrons below the Fermi surface as a result of the topological properties induced by the SO coupling in Bloch bands and consequently, the longitudinal conductance is zero. When the Fermi energy crosses the conduction band $\epsilon_{F}>M$, extrinsic spin-orbit mechanisms including side-jump and skew scattering contribute to the anomalous Hall conductivity. For $\lambda_{0}<0$ ($\lambda_{0}>0$), an increase in the extrinsic SO scattering strength leads to an increase (decrease) in the transport time and an increase (decrease) in the anomalous Hall and longitudinal conductivity values. As can be observed from Fig.~\ref{fig.1}(a), increasing the Fermi energy for a fixed magnetization (i.e., $M=10$ meV), the anomalous Hall conductivity decreases. In this case, the longitudinal conductance increases because $\sigma_{xx}\sim{\epsilon_{F}}$ as shown in the inset of Fig.~\ref{fig.1}(a). We plot the various contributions to the anomalous Hall conductivity for a fixed value of Fermi energy $\epsilon_{F}=200$ meV as seen in Figure \ref{fig.1}(b). In this figure, the extrinsic-scattering contribution dominates at larger values of the SO scattering strength.

\section{Conclusion}~\label{Con}

We have investigated the SOT and electronic AHE due to massless Dirac fermions in a 2D topological insulator with spin-momentum locking in the presence of a magnetization perpendicular to the topological insulator plane. We found that both scalar and extrinsic SO scattering play important roles in determining the magnitude of the spin-orbit torque. 

The Fermi energy, external magnetization strength, and extrinsic impurity scattering all affect the field-like and damping-like SOTs. The SOTs reach a maximum value by modifying the $M$, $\varepsilon _F$, and $\lambda$. Despite the fact that the strength of the damping-like SOT component is independent of the impurity density, the field-like SOT component decreases in a disordered system. The anomalous Hall conductivity, on the other hand, is independent of the impurity strength in the case of short-range impurity scattering, but highly nonlinear in the extrinsic SO scattering strength.

The physics for an in-plane magnetization is highly non-trivial and wholly different from that discussed here. It will be addressed in a future publication. 

\section{Acknowledgments}
This project is supported by Future Fellowship FT190100062.

\section{References}
\bibliographystyle{unsrt}
\bibliography{ref1}

\begin{thebibliography}{10}

\bibitem{PhysRevB.66.014407}
M.~D. Stiles and A.~Zangwill.
\newblock Anatomy of spin-transfer torque.
\newblock {\em Phys. Rev. B}, 66:014407, Jun 2002.

\bibitem{PhysRevB.80.134403}
Ion Garate and A.~H. MacDonald.
\newblock Influence of a transport current on magnetic anisotropy in gyrotropic
  ferromagnets.
\newblock {\em Phys. Rev. B}, 80:134403, Oct 2009.

\bibitem{PhysRevB.86.014416}
D.~A. Pesin and A.~H. MacDonald.
\newblock Quantum kinetic theory of current-induced torques in rashba
  ferromagnets.
\newblock {\em Phys. Rev. B}, 86:014416, Jul 2012.

\bibitem{RevModPhys.91.035004}
A.~Manchon, J.~\ifmmode~\check{Z}\else \v{Z}\fi{}elezn\'y, I.~M. Miron,
  T.~Jungwirth, J.~Sinova, A.~Thiaville, K.~Garello, and P.~Gambardella.
\newblock Current-induced spin-orbit torques in ferromagnetic and
  antiferromagnetic systems.
\newblock {\em Rev. Mod. Phys.}, 91:035004, Sep 2019.

\bibitem{chernyshov2009evidence}
Alexandr Chernyshov, Mason Overby, Xinyu Liu, Jacek~K Furdyna, Yuli
  Lyanda-Geller, and Leonid~P Rokhinson.
\newblock Evidence for reversible control of magnetization in a ferromagnetic
  material by means of spin--orbit magnetic field.
\newblock {\em Nature Physics}, 5(9):656--659, 2009.

\bibitem{ryu2013chiral}
Kwang-Su Ryu, Luc Thomas, See-Hun Yang, and Stuart Parkin.
\newblock Chiral spin torque at magnetic domain walls.
\newblock {\em Nature nanotechnology}, 8(7):527--533, 2013.

\bibitem{miron2011perpendicular}
Ioan~Mihai Miron, Kevin Garello, Gilles Gaudin, Pierre-Jean Zermatten, Marius~V
  Costache, St{\'e}phane Auffret, S{\'e}bastien Bandiera, Bernard Rodmacq,
  Alain Schuhl, and Pietro Gambardella.
\newblock Perpendicular switching of a single ferromagnetic layer induced by
  in-plane current injection.
\newblock {\em Nature}, 476(7359):189--193, 2011.

\bibitem{wadley2016electrical}
Peter Wadley, Bryn Howells, J~{\v{Z}}elezn{\`y}, Carl Andrews, Victoria Hills,
  Richard~P Campion, Vit Nov{\'a}k, K~Olejn{\'\i}k, F~Maccherozzi, SS~Dhesi,
  et~al.
\newblock Electrical switching of an antiferromagnet.
\newblock {\em Science}, 351(6273):587--590, 2016.

\bibitem{bandyopadhyay2008introduction}
Supriyo Bandyopadhyay and Marc Cahay.
\newblock {\em Introduction to spintronics}.
\newblock CRC press, 2008.

\bibitem{Ramaswamy2018}
Rajagopalan Ramaswamy, Jong~Min Lee, Kaiming Cai, and Hyunsoo Yang.
\newblock {Recent advances in spin-orbit torques: Moving towards device
  applications}.
\newblock {\em Applied Physics Reviews}, 5(3):031107, 9 2018.

\bibitem{Shao2021RoadmapOS}
Qiming Shao, Peng Li, Luqiao Liu, Hyunsoo Yang, Shunsuke Fukami, Armin Razavi,
  Hao Wu, Kang Wang, Frank Freimuth, Yuriy Mokrousov, Mark~D. Stiles, Satoru
  Emori, Axel Hoffmann, Johan Akerman, Kaushik Roy, Jian-Ping Wang, See-Hun
  Yang, Kevin Garello, and Wei Zhang.
\newblock Roadmap of spin–orbit torques.
\newblock {\em IEEE Transactions on Magnetics}, 57(7):1–39, 7 2021.

\bibitem{Nikolicnikolic2018}
Branislav~K Nikoli{\'c}, Kapildeb Dolui, Marko~D Petrovi{\'c}, Petr
  Plech{\'a}{\v{c}}, Troels Markussen, and Kurt Stokbro.
\newblock First-principles quantum transport modeling of spin-transfer and
  spin-orbit torques in magnetic multilayers.
\newblock {\em Handbook of Materials Modeling: Applications: Current and
  Emerging Materials}, pages 499--533, 2020.

\bibitem{zhou2021modulation}
Jing Zhou, Xinyu Shu, Weinan Lin, Ding~Fu Shao, Shaohai Chen, Liang Liu, Ping
  Yang, Evgeny~Y Tsymbal, and Jingsheng Chen.
\newblock Modulation of spin--orbit torque from srruo3 by
  epitaxial-strain-induced octahedral rotation.
\newblock {\em Advanced Materials}, 33(30):2007114, 2021.

\bibitem{RevModPhys.76.323}
Igor \ifmmode \check{Z}\else \v{Z}\fi{}uti\ifmmode~\acute{c}\else \'{c}\fi{},
  Jaroslav Fabian, and S.~Das~Sarma.
\newblock Spintronics: Fundamentals and applications.
\newblock {\em Rev. Mod. Phys.}, 76:323--410, Apr 2004.

\bibitem{RevModPhys.87.1213}
Jairo Sinova, Sergio~O. Valenzuela, J.~Wunderlich, C.~H. Back, and
  T.~Jungwirth.
\newblock Spin hall effects.
\newblock {\em Rev. Mod. Phys.}, 87:1213--1260, Oct 2015.

\bibitem{dieny2020opportunities}
Bernard Dieny, Ioan~Lucian Prejbeanu, Kevin Garello, Pietro Gambardella, Paulo
  Freitas, Ronald Lehndorff, Wolfgang Raberg, Ursula Ebels, Sergej~O
  Demokritov, Johan Akerman, et~al.
\newblock Opportunities and challenges for spintronics in the microelectronics
  industry.
\newblock {\em Nature Electronics}, 3(8):446--459, 2020.

\bibitem{shi2019all}
Shuyuan Shi, Shiheng Liang, Zhifeng Zhu, Kaiming Cai, Shawn~D Pollard, Yi~Wang,
  Junyong Wang, Qisheng Wang, Pan He, Jiawei Yu, et~al.
\newblock All-electric magnetization switching and dzyaloshinskii--moriya
  interaction in wte2/ferromagnet heterostructures.
\newblock {\em Nature nanotechnology}, 14(10):945--949, 2019.

\bibitem{macneill2017control}
D~MacNeill, GM~Stiehl, MHD Guimaraes, RA~Buhrman, J~Park, and DC~Ralph.
\newblock Control of spin--orbit torques through crystal symmetry in
  wte2/ferromagnet bilayers.
\newblock {\em Nature Physics}, 13(3):300--305, 2017.

\bibitem{kent2015new}
Andrew~D Kent and Daniel~C Worledge.
\newblock A new spin on magnetic memories.
\newblock {\em Nature nanotechnology}, 10(3):187--191, 2015.

\bibitem{PhysRevB.95.094401}
I.~A. Ado, Oleg~A. Tretiakov, and M.~Titov.
\newblock Microscopic theory of spin-orbit torques in two dimensions.
\newblock {\em Phys. Rev. B}, 95:094401, Mar 2017.

\bibitem{PhysRevB.96.045428}
Cong Xiao and Qian Niu.
\newblock Semiclassical theory of spin-orbit torques in disordered multiband
  electron systems.
\newblock {\em Phys. Rev. B}, 96:045428, Jul 2017.

\bibitem{Ralph2008}
D.C. Ralph and M.D. Stiles.
\newblock Spin transfer torques.
\newblock {\em Journal of Magnetism and Magnetic Materials},
  320(7):1190–1216, Apr 2008.

\bibitem{rozhansky2021asymmetric}
Igor Rozhansky, Konstantin Denisov, Mikhail Rakitskii, Nikita Averkiev, Henri
  Jaffres, and Henri-Jean Drouhin.
\newblock Asymmetric scattering and tunneling of electrons due to spin-orbit
  and exchange interaction.
\newblock In {\em Spintronics XIV}, volume 11805, page 1180514. SPIE, 2021.

\bibitem{mellnik2014spin}
AR~Mellnik, JS~Lee, A~Richardella, JL~Grab, PJ~Mintun, Mark~H Fischer,
  Abolhassan Vaezi, Aurelien Manchon, E-A Kim, Nitin Samarth, et~al.
\newblock Spin-transfer torque generated by a topological insulator.
\newblock {\em Nature}, 511(7510):449--451, 2014.

\bibitem{fan2014magnetization}
Yabin Fan, Pramey Upadhyaya, Xufeng Kou, Murong Lang, So~Takei, Zhenxing Wang,
  Jianshi Tang, Liang He, Li-Te Chang, Mohammad Montazeri, et~al.
\newblock Magnetization switching through giant spin--orbit torque in a
  magnetically doped topological insulator heterostructure.
\newblock {\em Nature materials}, 13(7):699--704, 2014.

\bibitem{PhysRevLett.114.257202}
Yi~Wang, Praveen Deorani, Karan Banerjee, Nikesh Koirala, Matthew Brahlek,
  Seongshik Oh, and Hyunsoo Yang.
\newblock Topological surface states originated spin-orbit torques in
  ${\mathrm{bi}}_{2}{\mathrm{se}}_{3}$.
\newblock {\em Phys. Rev. Lett.}, 114:257202, Jun 2015.

\bibitem{PhysRevLett.119.077702}
Jiahao Han, A.~Richardella, Saima~A. Siddiqui, Joseph Finley, N.~Samarth, and
  Luqiao Liu.
\newblock Room-temperature spin-orbit torque switching induced by a topological
  insulator.
\newblock {\em Phys. Rev. Lett.}, 119:077702, Aug 2017.

\bibitem{Fan2014}
Yabin Fan, Pramey Upadhyaya, Xufeng Kou, Murong Lang, So~Takei, Zhenxing Wang,
  Jianshi Tang, Liang He, Li-Te Chang, Mohammad Montazeri, Guoqiang Yu, Wanjun
  Jiang, Tianxiao Nie, Robert~N Schwartz, Yaroslav Tserkovnyak, and Kang~L
  Wang.
\newblock {Magnetization switching through giant spin-orbit torque in a
  magnetically doped topological insulator heterostructure}.
\newblock {\em Nature Materials}, 13:699--704, 2014.

\bibitem{Wang2017}
Yi~Wang, Dapeng Zhu, Yang Wu, Yumeng Yang, Jiawei Yu, Rajagopalan Ramaswamy,
  Rahul Mishra, Shuyuan Shi, Mehrdad Elyasi, Kie-Leong Teo, Yihong Wu, and
  Hyunsoo Yang.
\newblock {Room temperature magnetization switching in topological
  insulator-ferromagnet heterostructures by spin-orbit torques}.
\newblock {\em Nature Communications}, 8:1364, 2017.

\bibitem{Li2019}
Peng Li, James Kally, Steven S.-L. Zhang, Timothy Pillsbury, Jinjun Ding,
  Gyorgy Csaba, Junjia Ding, J.~S. Jiang, Yunzhi Liu, Robert Sinclair, Chong
  Bi, August DeMann, Gaurab Rimal, Wei Zhang, Stuart~B. Field, Jinke Tang,
  Weigang Wang, Olle~G. Heinonen, Valentine Novosad, Axel Hoffmann, Nitin
  Samarth, and Mingzhong Wu.
\newblock Magnetization switching using topological surface states.
\newblock {\em Science Advances}, 5(8):3415, 2019.

\bibitem{liu2021}
Xiaoyang Liu, Di~Wu, Liyang Liao, Peng Chen, Yong Zhang, Fenghua Xue, Qi~Yao,
  Cheng Song, Kang~L Wang, and Xufeng Kou.
\newblock Temperature dependence of spin—orbit torque-driven magnetization
  switching in in situ grown bi2te3/mnte heterostructures.
\newblock {\em Applied Physics Letters}, 118(11):112406, 2021.

\bibitem{fukami2016magnetization}
Shunsuke Fukami, Chaoliang Zhang, Samik DuttaGupta, Aleksandr Kurenkov, and
  Hideo Ohno.
\newblock Magnetization switching by spin--orbit torque in an
  antiferromagnet--ferromagnet bilayer system.
\newblock {\em Nature materials}, 15(5):535--541, 2016.

\bibitem{culcer2009}
Dimitrie Culcer, ME~Lucassen, RA~Duine, and R~Winkler.
\newblock Current-induced spin torques in iii-v ferromagnetic semiconductors.
\newblock {\em Physical Review B}, 79(15):155208, 2009.

\bibitem{Ramaswamy2019}
Rajagopalan Ramaswamy, Tanmay Dutta, Shiheng Liang, Guang Yang, MSM Saifullah,
  and Hyunsoo Yang.
\newblock Spin orbit torque driven magnetization switching with sputtered
  bi2se3 spin current source.
\newblock {\em Journal of Physics D: Applied Physics}, 52(22):224001, 2019.

\bibitem{pan2022efficient}
Quanjun Pan, Yuting Liu, Hao Wu, Peng Zhang, Hanshen Huang, Christopher
  Eckberg, Xiaoyu Che, Yingying Wu, Bingqian Dai, Qiming Shao, et~al.
\newblock Efficient spin-orbit torque switching of perpendicular magnetization
  using topological insulators with high thermal tolerance.
\newblock {\em Advanced Electronic Materials}, page 2200003, 2022.

\bibitem{PhysRevLett.113.196601}
Y.~Shiomi, K.~Nomura, Y.~Kajiwara, K.~Eto, M.~Novak, Kouji Segawa, Yoichi Ando,
  and E.~Saitoh.
\newblock Spin-electricity conversion induced by spin injection into
  topological insulators.
\newblock {\em Phys. Rev. Lett.}, 113:196601, Nov 2014.

\bibitem{PhysRevLett.116.096602}
J.-C. Rojas-S\'anchez, S.~Oyarz\'un, Y.~Fu, A.~Marty, C.~Vergnaud,
  S.~Gambarelli, L.~Vila, M.~Jamet, Y.~Ohtsubo, A.~Taleb-Ibrahimi,
  P.~Le~F\`evre, F.~Bertran, N.~Reyren, J.-M. George, and A.~Fert.
\newblock Spin to charge conversion at room temperature by spin pumping into a
  new type of topological insulator: $\ensuremath{\alpha}$-sn films.
\newblock {\em Phys. Rev. Lett.}, 116:096602, Mar 2016.

\bibitem{Ghosh2018}
S.~Ghosh and A.~Manchon.
\newblock {Spin-orbit torque in a three-dimensional topological
  insulator-ferromagnet heterostructure: Crossover between bulk and surface
  transport}.
\newblock {\em Physical Review B}, 97(13):134402, 4 2018.

\bibitem{cullen_2022}
James~H. Cullen, Rhonald~Burgos Atencia, and Dimitrie Culcer.
\newblock Electrically-induced spin torques due to the bulk states of
  topological insulators.
\newblock {\em arXiv:2206.09939}, 2022.

\bibitem{edelstein1990spin}
Victor~M Edelstein.
\newblock Spin polarization of conduction electrons induced by electric current
  in two-dimensional asymmetric electron systems.
\newblock {\em Solid State Communications}, 73(3):233--235, 1990.

\bibitem{mihai2010current}
Ioan Mihai~Miron, Gilles Gaudin, St{\'e}phane Auffret, Bernard Rodmacq, Alain
  Schuhl, Stefania Pizzini, Jan Vogel, and Pietro Gambardella.
\newblock Current-driven spin torque induced by the rashba effect in a
  ferromagnetic metal layer.
\newblock {\em Nature materials}, 9(3):230--234, 2010.

\bibitem{Dc2018}
Mahendra Dc, Roberto Grassi, Jun-Yang Chen, Mahdi Jamali, Danielle~Reifsnyder
  Hickey, Delin Zhang, Zhengyang Zhao, Hongshi Li, P~Quarterman, Yang Lv,
  Mo~Li, Aurelien Manchon, K~Andre Mkhoyan, Tony Low, and Jian-Ping Wang.
\newblock {Room-temperature high spin-orbit torque due to quantum confinement
  in sputtered Bi x Se (1-x) films}.
\newblock {\em Nature Materials}, 17:800--817, 2018.

\bibitem{Kurebayashi2019}
Daichi Kurebayashi and Naoto Nagaosa.
\newblock Theory of current-driven dynamics of spin textures on a surface of
  topological insulators.
\newblock {\em Physical Review B}, 100:134407, 7 2019.

\bibitem{siu2018}
Zhuo~Bin Siu, Yi~Wang, Hyunsoo Yang, and Mansoor~BA Jalil.
\newblock Spin accumulation in topological insulator thin films—influence of
  bulk and topological surface states.
\newblock {\em Journal of Physics D: Applied Physics}, 51(42):425301, 2018.

\bibitem{bonell2020}
Fr{\'e}d{\'e}ric Bonell, Minori Goto, Guillaume Sauthier, Juan~F Sierra,
  Adriana~I Figueroa, Marius~V Costache, Shinji Miwa, Yoshishige Suzuki, and
  Sergio~O Valenzuela.
\newblock Control of spin--orbit torques by interface engineering in
  topological insulator heterostructures.
\newblock {\em Nano Letters}, 20(8):5893--5899, 2020.

\bibitem{bhattacharjee2022effects}
Nirjhar Bhattacharjee, Krishnamurthy Mahalingam, Adrian Fedorko, Alexandria
  Will-Cole, Jaehyeon Ryu, Michael Page, Michael McConney, Hui Fang, Don
  Heiman, and Nian~Xiang Sun.
\newblock Effects of crystalline disorder on interfacial and magnetic
  properties of sputtered topological insulator/ferromagnet heterostructures.
\newblock {\em arXiv preprint arXiv:2205.09913}, 2022.

\bibitem{d1971possibility}
Mikhail~I D'Yakonov and VI~Perel.
\newblock Possibility of orienting electron spins with current.
\newblock {\em ZhETF Pisma Redaktsiiu}, 13:657, 1971.

\bibitem{PhysRevLett.83.1834}
J.~E. Hirsch.
\newblock Spin hall effect.
\newblock {\em Phys. Rev. Lett.}, 83:1834--1837, Aug 1999.

\bibitem{PhysRevLett.95.166605}
Hans-Andreas Engel, Bertrand~I. Halperin, and Emmanuel~I. Rashba.
\newblock Theory of spin hall conductivity in $n$-doped gaas.
\newblock {\em Phys. Rev. Lett.}, 95:166605, Oct 2005.

\bibitem{PhysRevB.2.4559}
L.~Berger.
\newblock Side-jump mechanism for the hall effect of ferromagnets.
\newblock {\em Phys. Rev. B}, 2:4559--4566, Dec 1970.

\bibitem{RevModPhys.82.3045}
M.~Z. Hasan and C.~L. Kane.
\newblock Colloquium: Topological insulators.
\newblock {\em Rev. Mod. Phys.}, 82:3045--3067, Nov 2010.

\bibitem{RevModPhys.83.1057}
Xiao-Liang Qi and Shou-Cheng Zhang.
\newblock Topological insulators and superconductors.
\newblock {\em Rev. Mod. Phys.}, 83:1057--1110, Oct 2011.

\bibitem{moore2010birth}
Joel~E Moore.
\newblock The birth of topological insulators.
\newblock {\em Nature}, 464(7286):194--198, 2010.

\bibitem{shen2012topological}
Shun-Qing Shen.
\newblock {\em Topological insulators}, volume 174.
\newblock Springer, 2012.

\bibitem{franz2013topological}
Marcel Franz and Laurens Molenkamp.
\newblock {\em Topological insulators}.
\newblock Elsevier, 2013.

\bibitem{Chang2015}
Po~Hao Chang, Troels Markussen, S{\o}ren Smidstrup, Kurt Stokbro, and
  Branislav~K. Nikoli{\'{c}}.
\newblock {Nonequilibrium spin texture within a thin layer below the surface of
  current-carrying topological insulator Bi2Se3: A first-principles quantum
  transport study}.
\newblock {\em Physical Review B - Condensed Matter and Materials Physics},
  92(20):201406, 11 2015.

\bibitem{Vobornik2011}
Ivana Vobornik, Unnikrishnan Manju, Jun Fujii, Francesco Borgatti, Piero
  Torelli, Damjan Krizmancic, Yew~San Hor, Robert~J. Cava, and Giancarlo
  Panaccione.
\newblock Magnetic proximity effect as a pathway to spintronic applications of
  topological insulators.
\newblock {\em Nano Letters}, 11:4079--4082, 2011.

\bibitem{PhysRevResearch.4.013001}
Rhonald~Burgos Atencia, Qian Niu, and Dimitrie Culcer.
\newblock Semiclassical response of disordered conductors: Extrinsic carrier
  velocity and spin and field-corrected collision integral.
\newblock {\em Phys. Rev. Research}, 4:013001, Jan 2022.

\bibitem{PhysRevLett.93.206602}
F.~D.~M. Haldane.
\newblock Berry curvature on the fermi surface: Anomalous hall effect as a
  topological fermi-liquid property.
\newblock {\em Phys. Rev. Lett.}, 93:206602, Nov 2004.

\bibitem{RevModPhys.82.1539}
Naoto Nagaosa, Jairo Sinova, Shigeki Onoda, A.~H. MacDonald, and N.~P. Ong.
\newblock Anomalous hall effect.
\newblock {\em Rev. Mod. Phys.}, 82:1539--1592, May 2010.

\bibitem{Liu_AnnRev2016}
Chao-Xing Liu, Shou-Cheng Zhang, and Xiao-Liang Qi.
\newblock The quantum anomalous hall effect: Theory and experiment.
\newblock {\em Annual Review of Condensed Matter Physics}, 7(1):301--321, 2016.

\bibitem{PhysRevLett.88.207208}
T.~Jungwirth, Qian Niu, and A.~H. MacDonald.
\newblock Anomalous hall effect in ferromagnetic semiconductors.
\newblock {\em Phys. Rev. Lett.}, 88:207208, May 2002.

\bibitem{Yu_2010}
R.~Yu, W.~Zhang, H.-J. Zhang, S.-C. Zhang, X.~Dai, and Z.~Fang.
\newblock Quantized anomalous hall effect in magnetic topological insulators.
\newblock {\em Science}, 329(5987):61--64, jun 2010.

\bibitem{Burkov_2014}
A.~A. Burkov.
\newblock Anomalous hall effect in weyl metals.
\newblock {\em Phys. Rev. Lett.}, 113:187202, Oct 2014.

\bibitem{PhysRevLett.104.146802}
Ion Garate and M.~Franz.
\newblock Inverse spin-galvanic effect in the interface between a topological
  insulator and a ferromagnet.
\newblock {\em Phys. Rev. Lett.}, 104:146802, Apr 2010.

\bibitem{PhysRevB.81.241410}
Takehito Yokoyama, Jiadong Zang, and Naoto Nagaosa.
\newblock Theoretical study of the dynamics of magnetization on the topological
  surface.
\newblock {\em Phys. Rev. B}, 81:241410, Jun 2010.

\bibitem{PhysRevB.96.014408}
Papa~B. Ndiaye, C.~A. Akosa, M.~H. Fischer, A.~Vaezi, E.-A. Kim, and
  A.~Manchon.
\newblock Dirac spin-orbit torques and charge pumping at the surface of
  topological insulators.
\newblock {\em Phys. Rev. B}, 96:014408, Jul 2017.

\bibitem{Liu2010}
Chao-Xing Liu, Xiao-Liang Qi, HaiJun Zhang, Xi~Dai, Zhong Fang, and Shou-Cheng
  Zhang.
\newblock Model hamiltonian for topological insulators.
\newblock {\em Phys. Rev. B}, 82:045122, 7 2010.

\bibitem{Hasan2010}
M.~Z. Hasan and C.~L. Kane.
\newblock Colloquium: Topological insulators.
\newblock {\em Rev. Mod. Phys.}, 82:3045--3067, 11 2010.

\bibitem{sinitsyn2007semiclassical}
NA~Sinitsyn.
\newblock Semiclassical theories of the anomalous hall effect.
\newblock {\em Journal of Physics: Condensed Matter}, 20(2):023201, 2007.

\bibitem{vasko2006quantum}
Fedir~T Vasko and Oleg~E Raichev.
\newblock {\em Quantum Kinetic Theory and Applications: Electrons, Photons,
  Phonons}.
\newblock Springer Science \& Business Media, 2006.

\bibitem{Culcer_2017}
Dimitrie Culcer, Akihiko Sekine, and Allan~H. MacDonald.
\newblock Interband coherence response to electric fields in crystals:
  Berry-phase contributions and disorder effects.
\newblock {\em Physical Review B}, 96(3), jul 2017.

\bibitem{Sekine_2017}
Akihiko Sekine, Dimitrie Culcer, and Allan~H. MacDonald.
\newblock Quantum kinetic theory of the chiral anomaly.
\newblock {\em Physical Review B}, 96(23), dec 2017.

\bibitem{PhysRevB.84.155440}
C.~M. Wang and F.~J. Yu.
\newblock Effects of hexagonal warping on surface transport in topological
  insulators.
\newblock {\em Phys. Rev. B}, 84:155440, Oct 2011.

\bibitem{liu2017weak}
Weizhe~Edward Liu, Ewelina~M Hankiewicz, and Dimitrie Culcer.
\newblock Weak localization and antilocalization in topological materials with
  impurity spin-orbit interactions.
\newblock {\em Materials}, 10(7):807, 2017.

\end{thebibliography}
\end{document}